\documentclass[prl,showpacs,amsmath,amssymb,twocolumn]{revtex4}
\usepackage{graphicx}
\usepackage[abs]{overpic}

\begin{document}

\newcommand{\be}{\begin{equation}}
\newcommand{\ee}{\end{equation}}
\newcommand{\ra}{\rangle}
\newcommand{\la}{\langle}

\title{Super-sharp resonances in chaotic wave scattering}

\author{Marcel Novaes}

\affiliation{Departamento de F\'isica, Universidade Federal de S\~ao
Carlos, S\~ao Carlos, SP, 13565-905, Brazil}

\begin{abstract}

Wave scattering in chaotic systems can be characterized by its spectrum of resonances, $z_n=E_n-i\frac{\Gamma_n}{2}$, where $E_n$ is related to the energy and
$\Gamma_n$ is the decay rate or width of the resonance.
If the corresponding ray dynamics is chaotic, a gap is believed to develop in the large-energy limit: almost all $\Gamma_n$ become larger than some $\gamma$. However, rare cases with
$\Gamma<\gamma$ may be present and actually dominate scattering
events. We consider the statistical properties of these super-sharp
resonances. We find that their number does not follow the fractal
Weyl law conjectured for the bulk of the spectrum. We also test, for a simple model, the universal predictions of random matrix theory for density of states inside the gap and the hereby derived probability distribution of gap size.

\end{abstract}


\maketitle

Scattering of waves in complex media is a vast area of research,
from oceanography and seismology through acoustics and optics, all
the way to the probability amplitude waves of quantum mechanics
\cite{book1,book3,book4,trends}. We focus our attention on the
important class of systems for which the complexity is not due to
the presence of randomness or impurities, but rather because the
corresponding ray dynamics is chaotic. The presence of multiple
scattering leads to very complicated cross-sections, with strongly
overlapping resonances that, although deterministic, have apparently
random positions and widths. It is not uncommon that in a given
situation only the sharpest resonances are relevant, with the others
providing an approximately uniform background.

For concreteness of terminology, we consider quantum mechanical
systems, but our results are general. We denote resonances by
$E-i\frac{\Gamma}{2}$ and call $\Gamma$ the width or the decay rate.
In the large-energy limit a gap develops in the resonance spectrum:
most resonances have their widths larger than $\gamma$, which is the
average decay rate of the corresponding classical (ray) dynamics.
This was noticed long ago \cite{gaspard}, and more recently there
have been attempts to prove it rigorously \cite{gap}. We are
interested in the rare case of states inside the gap, i.e. with
$\Gamma<\gamma$, which we call super-sharp resonances.

The distribution of typical resonances in chaotic systems is
conjectured to follow the so-called fractal Weyl law
\cite{prl91wl2003}: their number grows with $E$ as a power law, whose exponent is related to the fractal dimension $d$ of the classical repeller, the set of rays which remain trapped in the scattering region for infinite times,
both in the future and in the past \cite{gasp}. In a numerical experiment with a simple model, we find that the number of super-sharp resonances, denote it by $\mathcal{N}_{SSR}$, does not follow this law. It does grow with $E$ according to a power law, but the exponent seems to be insensitive to $\gamma$ or the dimension of the repeller.

We also investigate the dependence with energy of the width of the sharpest (and usually most important) resonance. Let this be denoted $\Gamma_0$. As $E$ grows, it is expected to converge to $\gamma$. More natural variables are their exponentials, and we find that $e^{-\Gamma_0}-e^{-\gamma}$ decreases with $E$ according to a power law, whose exponent is well approximated by $d-\alpha$.

A very fruitful approach to chaotic scattering of waves is random matrix theory (RMT) \cite{rmt}, in which the system's propagator (the Green's function of the wave
equation) is replaced by a random matrix, whose spectral properties
are studied statistically. The RMT prediction for the density of
resonance states inside the gap was derived in \cite{density}. In this work we derive the RMT prediction for the probability distribution of $e^{-\Gamma_0}-e^{-\gamma}$, and compare both these predictions to numerical results in a specific system.

Waves in chaotic systems can be modeled by the so-called `quantum
maps'. These are $N\times N$ unitary matrices where $N\sim 1/\hbar$, and the large energy limit $E\to\infty$ is replaced by the limit of large dimension, $N\to\infty$. This approach has
been used to study transport properties of semiconductor quantum
dots \cite{brouwer}, entanglement production \cite{entang}, the
fractal Weyl law \cite{weyl}, fractal wave functions
\cite{resonances,reso2}, proximity effects due to superconductors
\cite{bcs}, among other phenomena. Scattering can be introduced by
means of projectors. The propagator becomes a sub-unitary matrix of
dimension $M<N$, and its spectrum comprises $N-M$ zero eigenvalues
and $M$ resonances of the form $z_n=e^{-i(E_n-i\Gamma_n/2)}$.

As our dynamical model we use the kicked rotator. Its classical
dynamics is determined by canonical equations of motion in discrete
time,
\begin{align} q_{t+1}&=q_t+p_t+\frac{K}{4\pi}\sin(2\pi q_t)\quad ({\rm
mod}\;1)\\p_{t+1}&=p_t+\frac{K}{4\pi}[\sin(2\pi q_t)+\sin(2\pi
q_{t+1})]\quad ({\rm mod}\;1). \end{align} This system is known to
be fully chaotic for $K>7$, with Lyapunov exponent
$\lambda\approx\ln(K/2)$. Its quantization \cite{bcs,henning} yields a $N$ dimensional unitary matrix $U$. We set equal to zero a fraction of $1-\mu$ of its columns, corresponding to a `hole' in phase space. Here $\mu$ corresponds to the
fraction of rays which escape the scattering region per unit time,
so $\mu=e^{-\gamma}$. We do not consider very small values of $\mu$, which would correspond to widely open systems. 

\begin{figure}[t]
\includegraphics[angle=-90.0,scale=0.3,clip]{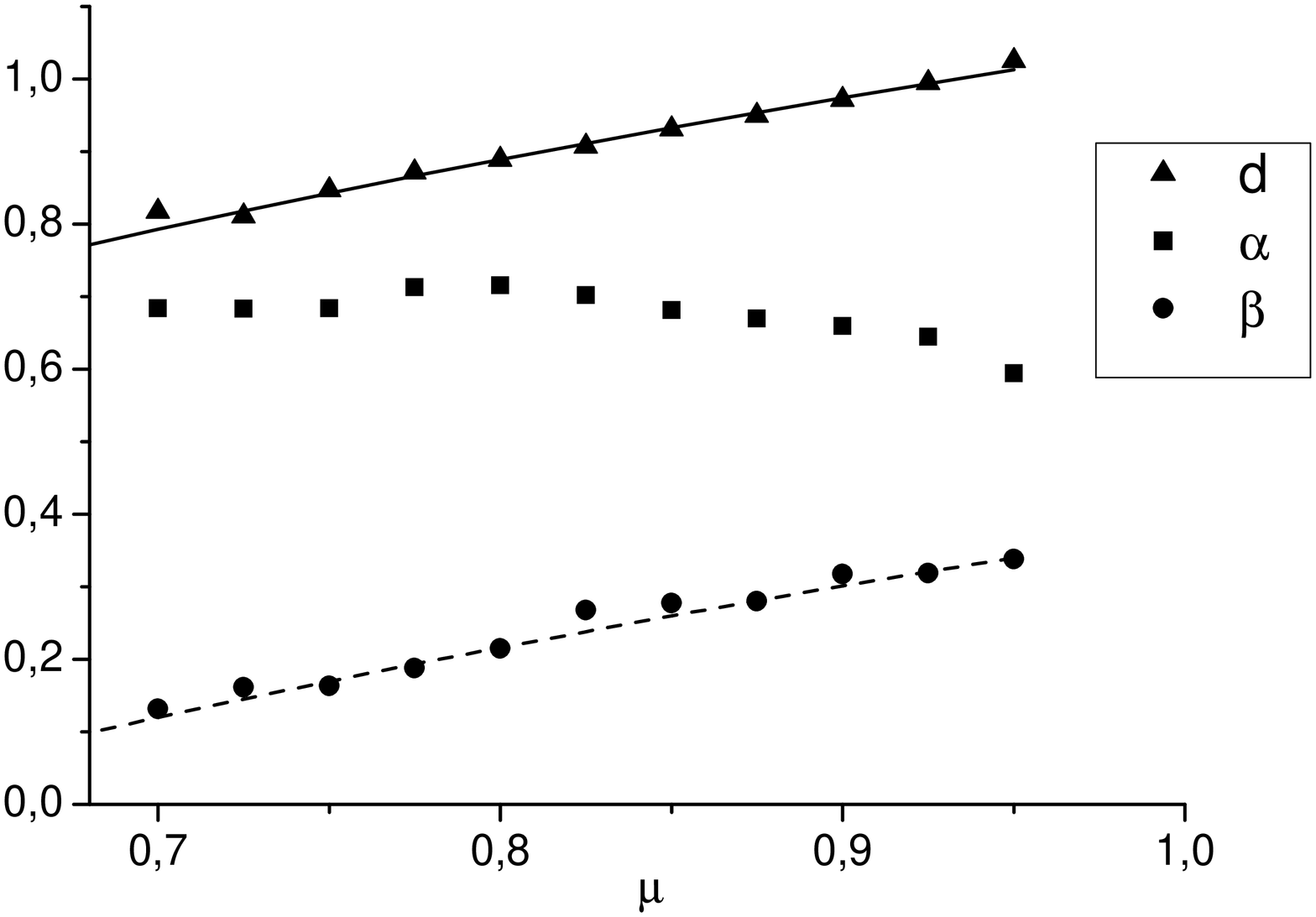}
\caption{Scaling exponents as functions of $\mu$ for the open kicked rotator. $d$, $\alpha$ and $\beta$ are related to the fractal Weyl law, the number of super-sharp resonances and the width of the gap, respectively. We see that $\alpha$ is approximately independent of the size of the opening. The solid line is a best linear fit to $d$, while the dashed line is $d-\langle\alpha\rangle$, where $\langle\alpha\rangle\approx 0.66$ is the average value of $\alpha$.}
\end{figure}

We find numerically that $\mathcal{N}_{SSR}\sim N^\alpha$, while  $e^{-\Gamma_0}-e^{-\gamma}\sim N^{-\beta}$. The exponents are plotted as functions of
$\mu$ in Figure 1, together with $d$, the numerically determined exponent in the Weyl law. We see that, somewhat surprisingly, in this range $\alpha$ is approximately constant, i.e. insensitive to the dimension of the repeller. The super-sharp resonances do not follow the fractal Weyl law. On the other hand, the exponent $\beta$ has approximately the same slope as $d$. We find the relation $d-\beta=\alpha$ to be approximately fulfilled, which is to be expected since it says that the number of super-sharp resonances, $\sim N^\alpha$, is proportional to the total number of resonances, $\sim N^d$, times the width of the gap, $\sim N^{-\beta}$.

\begin{figure}[t]
\includegraphics[angle=-90.0,scale=0.27 ]{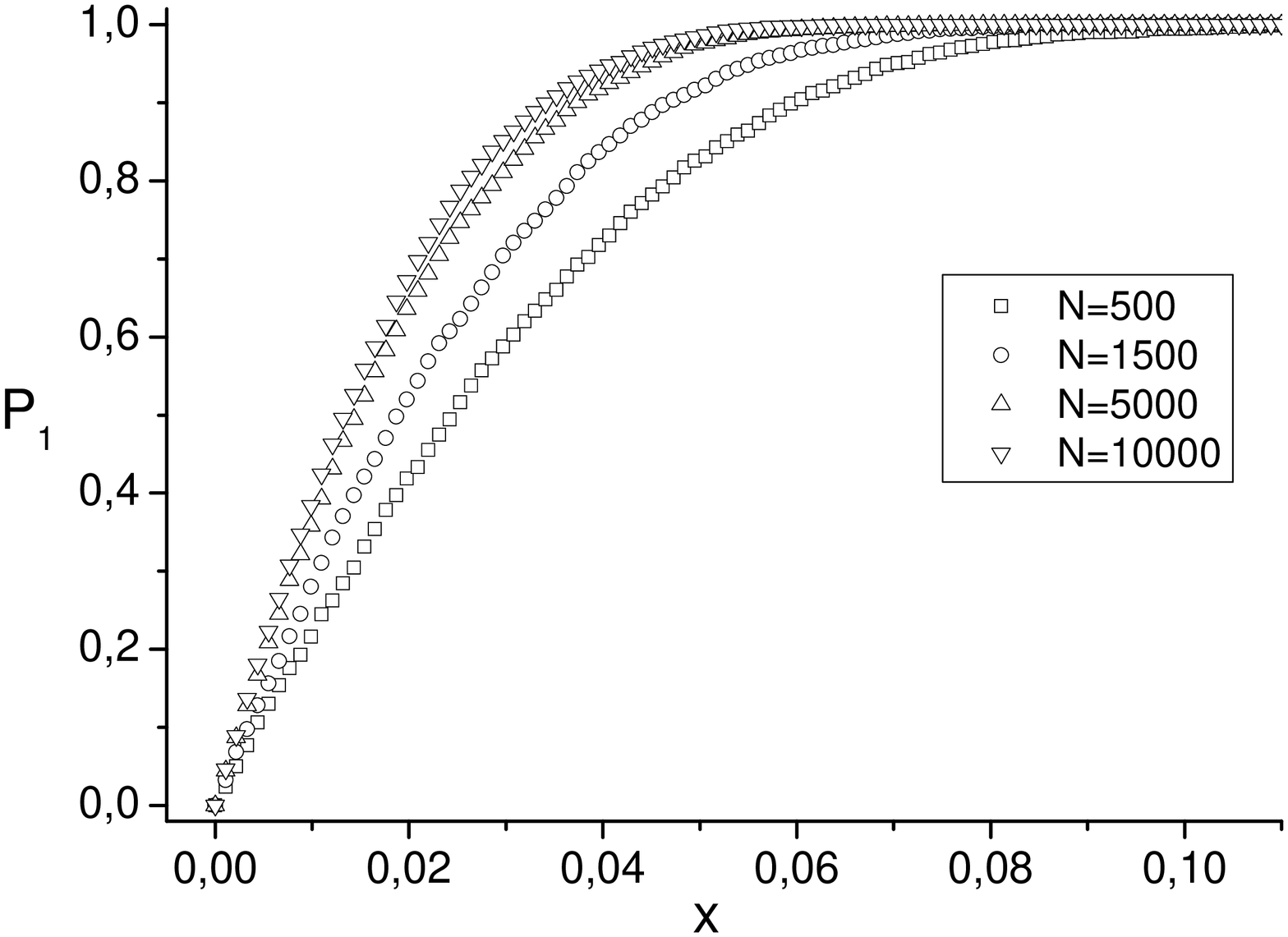}
\includegraphics[angle=-90.0,scale=0.27 ]{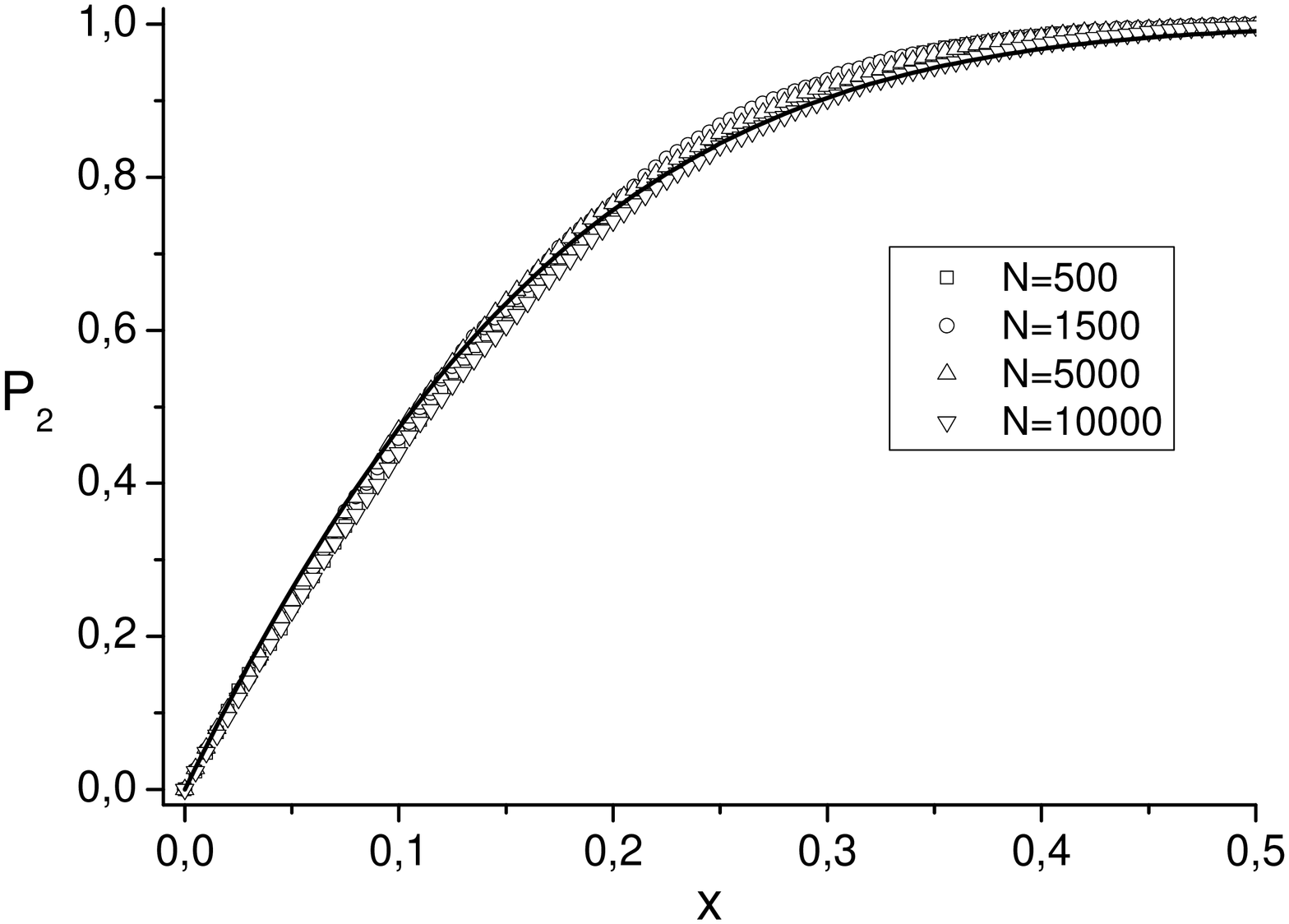}
\caption{Top panel: $P_1={\rm Pr}(r^2-\mu<x)$ for the open kicked rotator,
at several dimensions, for $\mu=0.8$. Lower panel: same for $P_2={\rm Pr}(\widetilde{\eta}(r^2-\mu)<x)$, where $\widetilde{\eta}$ is defined in the text. Solid line is a rescaled RMT prediction.}
\end{figure}

We now turn to an RMT treatment of the problem.
RMT for quantum maps amounts to taking matrices uniformly
distributed in the unitary group. In \cite{trunc}, an ensemble of
truncated unitary matrices was introduced as a model for scattering,
and it was shown that, as $N,M\to\infty$ with $\mu=M/N$ held fixed,
the probability density of $r=e^{-\Gamma/2}$ converges to \be
\label{Pr} P_\infty(r)=\frac{1-\mu}{\mu}\frac{2r}{(1-r^2)^2},\ee if
$r^2<\mu$ and to $0$ otherwise. The expression (\ref{Pr}) was tested
numerically for a chaotic quantum map in \cite{henning} and
found to be an accurate description of the bulk of the spectrum,
provided a rescaling was performed to incorporate the fractal Weyl
law.

Let us start with the density of super-sharp resonances. This calculation can be found in \cite{density}; we sketch it here for completeness. The probability density for $r=e^{-\Gamma/2}$ is known
exactly \cite{trunc}, and when
$M=\mu N$ and $N\gg 1$ it can be approximated to
\be \label{Pr2}P(r)\approx \frac{P_\infty(r)}{2}\left[1+{\rm
erf}\left(\eta\sqrt{\mu}\frac{(\mu-r^2)}{r}\right)\right],\ee
where erf is the error function and \be\eta=\sqrt{\frac{N}{2\mu(1-\mu)}}\ee is now the large number. Clearly, (\ref{Pr2}) will be different from
$P_\infty(r)$ only if $\mu-\nu_0$ is of the order of $1/\eta$. After setting
$r^2=\mu+\epsilon/\eta$, the probability distribution
for the variable $\epsilon$ becomes \be\label{Pe} \widetilde{P}(\epsilon)=\sqrt{\pi}(1-{\rm
erf}(\epsilon)). \ee This is the density of states inside the gap. It is only non-zero in a small region that shrinks as $N^{-1/2}$ in the asymptotic limit. Its integral provides the probability that $r^2-\mu$ be less than some value $x$. This we denote by Pr($r^2-\mu<x$).

In Figure 2a we see Pr($r^2-\mu<x$) for the open kicked rotator, at various dimensions for $\mu=0.8$. We see that as $N$ grows the values of $r$ become more localized around $\mu$. In Figure 2b we introduce a scaled variable $\widetilde{\eta}(r^2-\mu)$, but with $\widetilde{\eta}$ different from $\eta$ in that it involves the actual exponent $\beta$ instead of the RMT prediction $1/2$: \be \label{eta2} \widetilde{\eta}=\left(\frac{N}{2\mu(1-\mu)}\right)^\beta.\ee All curves fall on top of each other, indicating that this is the correct scaling. Moreover, the shape of the curve agrees with (\ref{Pe}).

Let us now consider the probability distribution of the
largest eigenvalue of the propagator, which corresponds to the
sharpest resonance and whose modulus we denote $R$. Largest eigenvalue distributions are important in different areas of mathematics \cite{largmath} and physics
\cite{largphys}, and have even been measured \cite{nir}. A similar calculation to the one below can be found in \cite{ginibre}. Let
$\mathcal{P}(\{z\})$ denote the joint probability density function
for all eigenvalues. When integrated over all variables from $0$ to
$x$, it gives the probability that all eigenvalues are smaller than
$x$. Therefore, if $Q(x)$ is the probability that the largest eigenvalue be smaller than
$x$, then \be {\rm Pr}(R<x)=
\int_{[0,x]^M}\mathcal{P}(\{z\})\prod_{i=1}^Md^2z_i. \ee

The jpdf of the eigenvalues is \cite{trunc}: \be
\mathcal{P}(\{z\})\propto \prod_{i<j}^{1..M}|z_i-z_j|^2
\prod_{i=1}^{M}(1-|z_i|^2)^{N-M-1}.\ee It is a usual trick to write
$\prod_{i<j}|z_i-z_j|^2$ in terms of the Vandermonde determinant
$|{\rm det}A|^2$, where $A_{ij}=z_j^{i-1}$. This can be shown to be
equal to $M!$det$B$, where $B_{ij}=z_i^{j-1}(z_i^*)^{i-1}$. Each
element of the matrix $B$ depends on a single variable, and the
integration decouples. The angular part of the integrals
diagonalizes the matrix and, if $M=\mu N$ and $N\gg1$, the result
becomes \be {\rm Pr}(R<x)\propto \prod_{j=0}^M
\int_0^x(1-y)^{N(1-\mu)}y^jdy. \ee This result is exact, but a bit
cumbersome. Approximating the integrand by a Gaussian
function we arrive at $ {\rm Pr}(R<x)\propto\prod_{\ell=0}^M\left\{
\frac{1}{2}+\frac{1}{2}{\rm
erf}\left[\eta\left(x^2-\mu+\frac{\ell(1-\mu)}{N}\right)\right]\right\}.$
This result can be further simplified by exponentiating the product
into a sum, setting the scale as $\ell=2\mu\eta\xi$ and
approximating the sum by an integral. We get \be\label{integ}
{\rm Pr}(R^2-\mu<x)\propto \exp\left\{2\mu\eta\int_0^\infty d\xi
\mathcal{L}\left(\eta x^2+\xi\right)\right\},\ee where we have
defined the function \be \mathcal{L}(z)=\ln\left(\frac{1+{\rm
erf}(z)}{2}\right).\ee

\begin{figure}[t]
\includegraphics[angle=-90.0,scale=0.27 ]{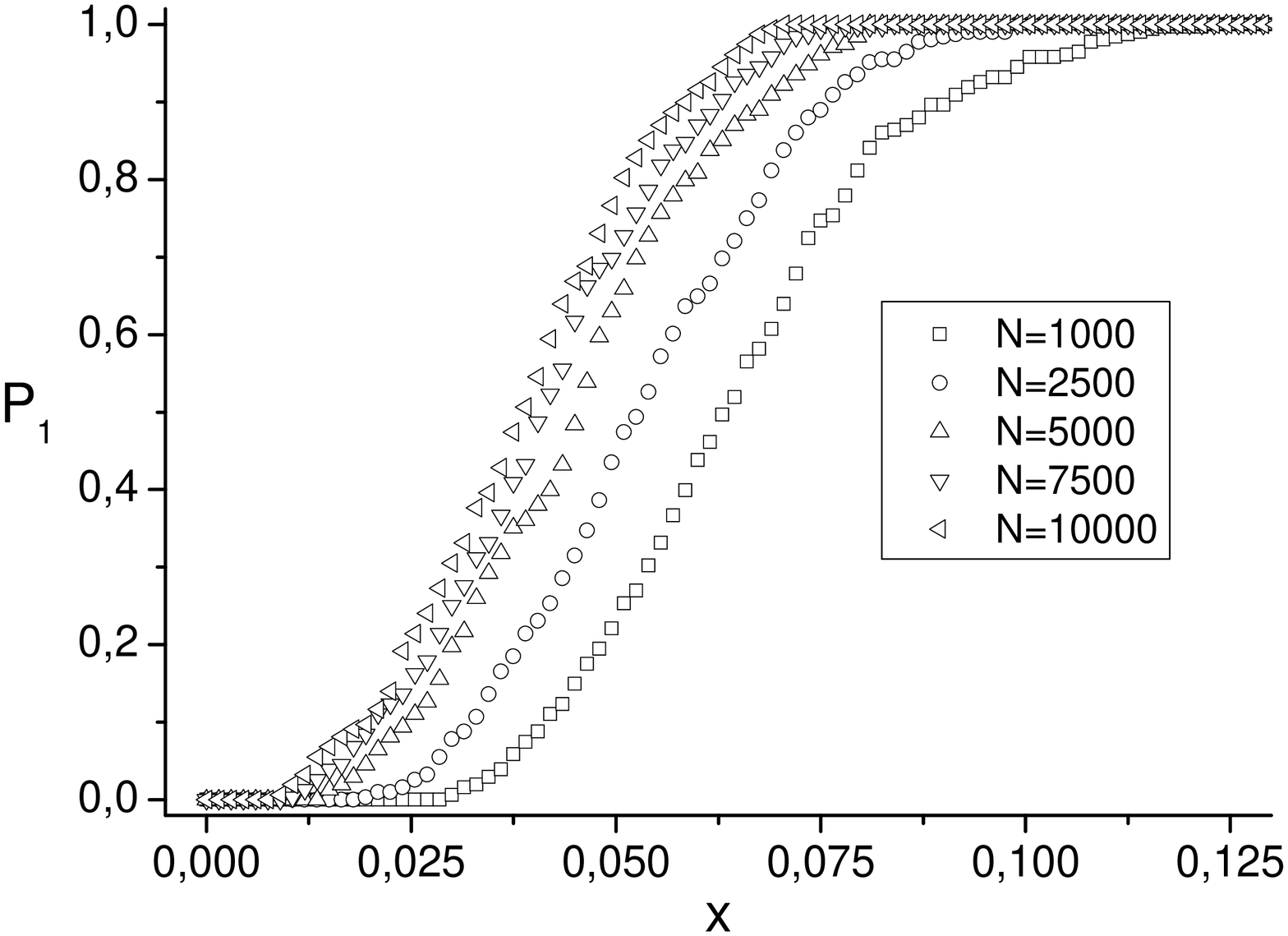}
\includegraphics[angle=-90.0,scale=0.27 ]{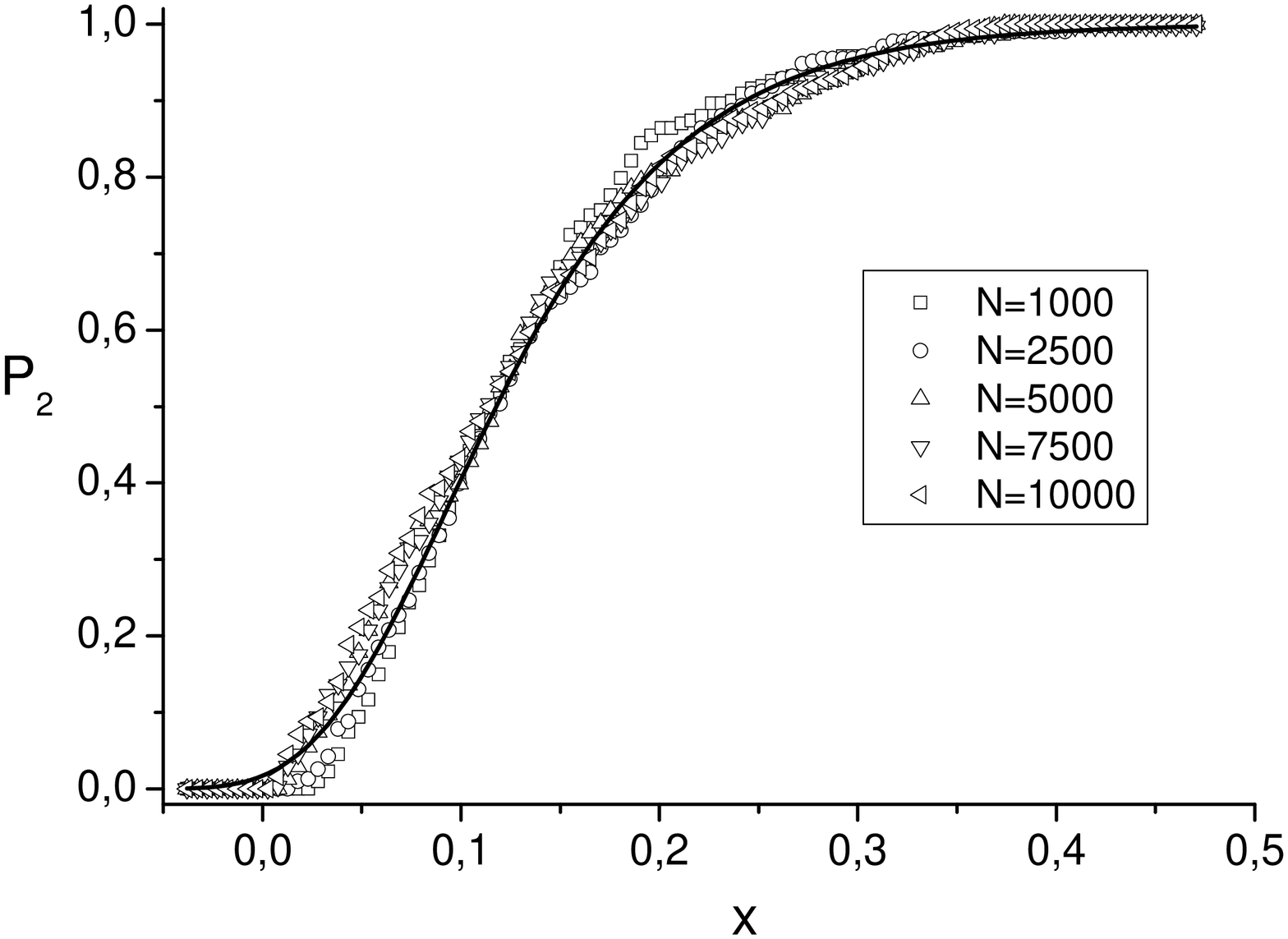}
\caption{Top panel: $P_1={\rm Pr}(R^2-\mu<x)$, where $R$ is the modulus of the largest eigenvalue, for the open kicked rotator at several dimensions, for $\mu=0.8$. Lower panel: same for $P_2={\rm Pr}(\widetilde{\eta}^2(R^2-\mu)^2<x)$. Solid line is a rescaled RMT prediction.}
\end{figure}

One interesting question that can be answered at this point is, how likely is it that a true gap will exist at $\mu$ for a finite value of $N$? The probability
that all eigenvalues are smaller than $\mu$ is simply given by the
exponential in (\ref{integ}) with $x=0$. It is thus proportional
to $e^{-c\sqrt{N}}$ for some constant $c$.

Notice that (\ref{integ}) has some similarity with the Tracy-Widom distribution \cite{tracy} of the largest eigenvalue of
Gaussian ensembles of RMT, in the sense that it involves the
exponential of the integral of a function that satisfies a
non-linear differential equation, $
\mathcal{L}''(z)=-2z\mathcal{L}'(z)
-\left(\mathcal{L}'(z)\right)^2.$ It is not a Painlev\'e transcendent, however.

Returning to the calculation, let us change variable to
$\delta=\eta x^2$ and obtain $\exp\left\{2\mu\eta\int_\delta^\infty
\mathcal{L}(z)dz\right\}.$ Clearly, this function does not converge as $\eta\to\infty$. Assuming $\delta$ to be large, we use $ \mathcal{L}(z)\approx -e^{-z^2}/2\sqrt{\pi}z$ and integrate by parts to get $ \exp\left\{-2\mu\eta e^{-\delta^2}/4\sqrt{\pi}\delta^2\right\}.$ In order to have a finite limit, we must have $\eta e^{-\delta^2}/\delta^2=O(1)$. This implies that $\delta^2=y+W{(\eta)}$, where $W$ is the Lambert function, which for large $\eta$ can be approximated as $W(\eta)\approx \log\eta-\log\log\eta$. Therefore, if instead of $R$ we consider the variable \be \label{log} \rho=\eta^2 (R^2-\mu)^2-\log\eta+\log\log\eta,\ee then \be\label{gumbel} {\rm Pr}(\rho<y)=\exp\{-\frac{\mu}{2\sqrt{\pi}}e^{-y}\},\ee which is a modified Gumbel function. Therefore, the distribution of the slightly awkward variable $\rho$ (see also \cite{ginibre}) has a limit as $\eta\to\infty$, but this limit is approached very slowly and finite $\eta$ calculations may present significant deviations.

In Figure 3a we see Pr($R^2-\mu<x$) for the open kicked rotator, at various dimensions for $\mu=0.8$. In Figure 3b we introduce the scaled variable $\widetilde{\eta}^2(R^2-\mu)^2$, where $\widetilde{\eta}$ is given by (\ref{eta2}). As a result, all curves fall on top of each other, indicating that this is the correct scaling. Notice that we must not introduce the $\log\widetilde{\eta}$ or $\log\log\widetilde{\eta}$ factors that appear in (\ref{log}), as they would spoil the scaling. The results agree very well with the function $\exp\{-e^{-ax+b}\}$, with fitted values of $a$ and $b$.

We close with some remarks on resonance eigenfunctions. These may be depicted in
phase space by means of their Husimi function, $H_\psi(q,p)=|\langle
q,p|\psi\rangle|^2$, where $|q,p\rangle$ is a coherent state. It was shown in \cite{resonances} that these Husimi functions are supported on the backward trapped
set, the unstable manifold of the repeller. How they are distributed
on this support depends on their decay rate: states with larger
$\Gamma$ concentrate on the dynamical pre-images of the opening,
while states with small $\Gamma$ concentrate on the repeller.
Semiclassically, \be\label{mass} \int_{R_m} H_{\psi_n}(q,p) dqdp
\approx |z_n|^{2m}(1-|z_n|^2),\ee where $R_m$ is the $m$-th preimage
of the opening. In principle, this relation would allow a state
whose decay rate equals the classical decay rate to be uniformly
distributed, because the area of $R_m$ decays like
$e^{-m\gamma}$. Therefore, super-sharp resonances must show an
increased degree of localization above uniformity. Indeed, since
they can also be seen as super-long-lived states, one would expect
them to be associated with periodic orbits (see for example
\cite{old,diego,periodicas,novo}). This topic deserves further
investigations. 

Another line of research to be followed would be to investigate a possible relation between the exponents $\alpha$ and $\beta$ to ray dynamics. In particular, it is not clear whether they depend on the Lyapounov exponent and whether they are universal or system specific.

This work was supported by CNPq and FAPESP.

\end{document}